\documentclass[12pt]{TD-CJS}

\usepackage{latexsym}
\usepackage{amsmath}
\usepackage{amsfonts}
\usepackage{amssymb}
\usepackage{psfrag}
\usepackage{graphicx}
\usepackage{xcolor}

\allowdisplaybreaks[1]

\begin{document}
%\firstpage{1}
%\lastpage{45}
%\jvol{xx}
%\issue{yy}
%\jyear{2018}
\jid{To appear in the Canadian Journal of Statistics.}
%\aid{???}
% The running head contains the author names
\rhauthor{Holbrook, Lumley and Gillen}
%\copyrightline{Statistical Society of Canada}
%\Frenchcopyrightline{Soci\'et\'e statistique du Canada}
% History: received and accepted dates
%\received{\rec{9}{July}{2009}}
%\accepted{\acc{8}{July}{2010}}

% User-defined commands go here
\renewcommand{\eqref}[1]{(\ref{#1})}
\newcommand{\mb}[1]{\mathbf{#1}}
\newcommand{\mbb}[1]{\mathbb{#1}}
\newcommand{\mt}[1]{\mathrm{#1}}
\newcommand{\rv}{random variable}
\newcommand{\var}{\mbox{\rm var}}
\newcommand{\Var}{\mbox{Var}}
\newcommand{\cov}{\mbox{\rm cov}}
\newcommand{\Cov}{\mbox{\rm Cov}}
\newcommand{\g}{\mathbf{g}}

% Title, authors, affiliations
\title[]{Estimating prediction error for complex samples}%\query{Q1}
 \author{Andrew Holbrook\authorref{1}}
 \author{Thomas Lumley\authorref{2}}
 \author{Daniel Gillen\authorref{1}}
  \affiliation[1]{Department of Statistics, University of California, Irvine, CA, U.S.A.}
 \affiliation[2]{Department of Statistics, University of Auckland, Auckland, New Zealand}

% Abstract, keywords, and classification codes
\startabstract{%
\keywords{
\KWDtitle{Key words and phrases}
AIC\sep Generalization error\sep Generalized linear models\sep  Horvitz-Thompson\sep NHANES III.
%  MSC 2010 subject classification codes
\KWDtitle{MSC 2010}Primary 62D05\sep secondary 62F40}
\begin{abstract}
\abstractsection{}
\abstractsection{Abstract}
With a growing interest in using non-representative samples to train prediction models for numerous outcomes it is necessary to account for the sampling design that gives rise to the data in order to assess the generalized predictive utility of a proposed prediction rule. After learning a prediction rule based on a non-uniform sample, it is of interest to estimate the rule's error rate when applied to unobserved members of the population. \cite{efron1986biased} proposed a general class of covariance penalty inflated prediction error estimators that assume the available training data are representative of the target population for which the prediction rule is to be applied. We extend Efron's estimator to the complex sample context by incorporating Horvitz-Thompson sampling weights and show that it is consistent for the true generalization error rate when applied to the underlying superpopulation. The resulting Horvitz-Thompson-Efron (HTE) estimator is equivalent to dAIC, a recent extension of AIC to survey sampling data, but is more widely applicable. The proposed methodology is assessed with simulations and is applied to models predicting renal function obtained from the large-scale NHANES survey.
%\abscopyright
%\fabstractsection{}
%\fabstract\Frenchabscopyrightsection{R\'{e}sum\'{e}}
%Ins\'{e}rer votre r\'{e}sum\'{e}
\end{abstract}}
\makechaptertitle

% Email address for corresponding author
 \correspondingauthor[*]{\\\email{aholbroo@uci.edu}}

\section{Introduction}

The goal of building prediction models using empirical samples has become ubiquitous throughout all areas of business and science.  With the exponential rise in statistical and machine learning methods for training flexible prediction models, increasing interest has been devoted to assessing the \emph{extra-sample} \citep{efron1997improvements} performance of candidate models when they are applied to unobserved members of the population of interest.  Analytic assessments of the performance of a prediction rule commonly focus on the expected loss associated with the rule, where the expectation is taken with respect to the underlying distribution of a new, independently sampled response conditional upon the observed support of the sampled predictors giving rise to the rule \citep{friedman2001elements}. It is widely recognized that computation of the loss function solely based on the training sample is optimistically biased for this expectation.

Statisticians have developed a number of techniques to adjust for this bias and thus ascertain extra-sample prediction performance. In this paper, we are interested in a class of unbiased estimators obtained by inflating the observed prediction error by a `covariance penalty'.  In linear regression under squared-error loss, Mallows's Cp \citep{efron1986biased} inflates the observed mean squared error by the covariance between each observed response and its fitted value. Relying on a similar covariance penalty, Stein's unbiased risk estimator is unbiased for mean squared error for differentiable prediction rules under Gaussian errors \citep{stein1981estimation}. Akaike's Information Criteria (AIC) \citep{akaike1998information} penalizes the observed negative log-likelihood (or deviance) by a term that is asymptotically equivalent to a covariance penalty and thus achieves an asymptotically unbiased prediction error estimate.   A more general treatment of covariance penalty inflated estimators for arbitrary prediction rules and loss functions has been considered by Efron \citep{efron1986biased,efron2004estimation}.

But how do we ascertain extra-sample prediction performance when the \emph{data} themselves are biased?  Ignoring complex, non-uniform, unbalanced or otherwise biased sampling will cause these `unbiased' estimators to be biased for future data arising from the true population of interest.  Policy makers, economists, statisticians, and health care professionals will choose sub-optimal prediction rules and make sub-optimal predictions.

Biased samples are common. Non-uniform random sampling designs are commonly employed throughout multiple empirical sciences because they afford researchers greater efficiency in estimating parameters specific to less prevalent sub-populations.  Classic examples of complex sampling designs include those implemented by the United States Census Bureau \citep{united2000current} and the National Health and Nutrition Examination Study (NHANES) \citep{nhanes3}. In each case, specific sub-populations are over-sampled by design and sampling weights are used to correct population prevalence estimates and draw inference for estimands at the population level.

Despite the growing interest in prediction modeling, there is a paucity of prediction error estimators for data arising from such complex sampling designs. \textit{This dearth is not for a lack of need.} Prediction models trained on health survey data are common and of interest within the medical and health-care communities. For example, \cite{wang2015forecasting} proposes statistical modeling methodologies for forecasting based on non-representative samples but does not propose any sort of model comparison criterion that would take sample complexity into account.    \cite{yu2010application} trains support vector machine models using the NHANES survey to predict diabetes (and boasts 140 citations). \cite{zhang2016prediction} uses NHANES to train a predictive model for peripheral arterial disease, and \cite{zhang2017prediction} does the same for predicting atherosclerotic cardiovascular disease mortality. There are many, many more such examples we do not list here.

We address this deficiency by extending Efron's covariance penalty inflated prediction error estimator to account for complex data.  The resulting estimator provides a unified framework for prediction model assessment that can be used for arbitrary loss functions and can be applied to regression based predictive models as well as algorithmically deduced prediction rules such as random forests  and k-nearest neighbour approaches. We establish consistency for the true error rate relative to the super-population and further show equivalence to dAIC \citep{lumley2015aic}, an extension of AIC for survey samples, as a special case in the context of  generalized linear regression models (GLM) \citep{mccullagh1989generalized}. 

\section{Prediction error estimation for simple random samples}\label{PeeSRS}

We begin with the classic problem of estimating the prediction error rate where data, ($y$, $X$), is obtained via a simple random sample with $y$ denoting the vector of outcomes of interest obtained on multiple sampling units and $X$ denoting a matrix of explanatory variables on sampling units. Assume that an unknown data generating mechanism defined by $\g$ has produced $y$, from which we estimate the expectation $\mu = E_{\g} (y)$ with $\hat{\mu} = m(y)$, where $m(\cdot)$ is an arbitrary function potentially obtained from the data. The in-sample error is given by
\begin{align}\label{eq::err1}
err \,\equiv \frac{1}{n} \sum_{i=1}^n err_i = \frac{1}{n} \sum_{i=1}^n Q\big(y_i, \hat{\mu}_i \big)  \, ,
\end{align}
where $n$ is the length of $y$, the number of observations, and $Q(\cdot,\cdot)$ denotes a specified loss function. Given the in-sample error, a common goal is to estimate
\begin{align}\label{eq::Err1}
Err \,\equiv \frac{1}{n}\sum_{i=1}^n Err_i= \frac{1}{n}\sum_{i=1}^n E_0 \, Q\big(y_i^0,\hat{\mu}_i\big)  
\end{align}
for fixed $\hat{\mu}_i$. Here $E_0$ denotes the expectation over an unobserved random variable $y_i^0$ drawn independently from mechanism $\g$ but conditioning on observed support $x_i$. Note that although $y_i^0$ shares the same covariates $x_i$ as observation $y_i$, the true data generating mechanism $\g$ may or may not be a function of observed covariates.  Typically, $err$ is a biased estimate for $Err$, but a correction is available by way of the covariance penalty. 

\subsection{The covariance penalty inflated estimator}\label{cov_inflate_est}

 Covariance penalty inflated prediction error estimators are unbiased for extra-sample prediction error.
We assume that the loss function $Q(\cdot,\cdot)$ belongs to the $q$-class of loss functions \citep{efron1986biased} . A member of the $q$-class of loss functions is constructed from some concave function $q(\cdot)$.  Given this function, the error for outcome $y_i$ and prediction $\hat{\mu}_i$ is given by
\begin{align*}
    Q(y_i,\hat{\mu}_i) = q(\hat{\mu}_i) + \dot{q}(\hat{\mu}_i)\, (y_i-\hat{\mu}_i) - q(y_i)\, ,
\end{align*}
where $\dot{q}(\cdot)$ is the derivative of $q(\cdot)$.  This is not a limiting assumption.  For example, the deviance functions of exponential family distributions belong to this class. Suppose $y_i$ follows an exponential family distribution
\begin{align*}
    g_{\mu_i} (y_i) = \exp \big( \lambda_i\, y_i - \psi(\lambda_i)\big) \, .
\end{align*}
Here, $\lambda_i$ is the natural parameter, $\psi$ enforces density function integration constraints, and $\mu_i$ and $\lambda_i$ are related by canonical link functions. Then the choice 
\begin{align}\label{q_func}
    q(y_i) = 2 \left(\psi (\hat{\lambda}_i) - y_i\, \hat{\lambda}_i \right)\, ,
\end{align}
with $\hat{\lambda}_i$ estimated from the observed data, renders the deviance \citep{mccullagh1989generalized} as in-sample error:
\begin{align}\label{deviance}
    err = \frac{1}{n} \sum_{i=1}^n Q\big(y_i, \hat{\mu}_i \big) = \frac{2}{n}\big(\log g_y(y) - \log g_{\hat{\mu}} (y) \big) \, ,
\end{align}
where $g_y(\cdot)$ and $g_{\hat{\mu}}(\cdot)$ are the exponential family likelihoods with mean parameters $y$ and $\hat{\mu}$, respectively.
Note that since only the second term in the deviance depends on  estimate $\hat{\mu}$, one may also consider concave function \eqref{q_func} as inducing the negative log-likelihood loss.
Since the in-sample error is often an underestimate of $Err$, we define the
\begin{align*}
    \textit{Optimism} \quad O_i = O_i(\g,y) = Err_i - err_i \, ,
\end{align*}
and the
\begin{align*}
    \textit{Expected optimism} \quad \Omega_i (\g) = E_\g \, O_i(\g,y) \, .
\end{align*}
If one is able to to obtain a consistent estimate of the optimism pertaining to a prediction rule, then the true generalization error may be estimated by adding the estimated optimism to the in-sample error $err$.
Within the $q$-class of loss functions, the optimism can be analytically estimated using the term
\begin{align*}
    \hat{\lambda}_i = -\dot{q}(\hat{\mu}_i)/2 \, .
\end{align*}
When $Q(y, \hat{\mu} )$ is the deviance for an exponential family distribution, $\hat{\lambda}_i$ is the estimated natural parameter for the $i$th observation \citep{efron1986biased,efron2004estimation}. Indeed, the following result identifies the expected optimism with the covariance between $y$ and $\hat{\boldmath{\lambda}}$.
\begin{theorem}{Theorem 2.1}{(The optimism theorem  \citep{efron1986biased,efron2004estimation})} For error measure $Q(y,\hat{\mu})$, we have
 \begin{align*}
     E_\g(  Err_i) = E_\g(err_i +  \Omega_i )\, ,
 \end{align*}
 where
 \begin{align}\label{optimism2}
     \Omega_i = 2\, \mbox{cov}_\g (y_i, \hat{\lambda}_i) \, .
 \end{align}
\end{theorem}\label{optimism}
It is a corollary \citep{efron1986biased} that when $\hat{\mu}$ is the MLE of $\mu$ for a correctly specified GLM, and when prediction error is given by the deviance, then
\begin{align}\label{cov_approx}
\frac{1}{n}\sum_{i=1}^n\Omega_i = E_\g \big(Err - err) \approx 2\, p /n 
\end{align}
for $p$ the number of model parameters.  This approximation is obtained through the Taylor expansion of the link function and is exact for the Gaussian case. Following Equation \eqref{cov_approx} the reader might not be surprised that the unbiased estimator arising from Formula \eqref{optimism2} is at least asymptotically equivalent to AIC \citep{efron1986biased}.  The same paper discusses the scenarios in which analytic estimators of the covariance penalty are available and how the parametric bootstrap may be used when they are not.

\section{Prediction error estimation for complex samples}\label{PeeCS}

While the covariance penalty inflated prediction error estimation procedure is useful for prediction rules derived from simple random samples, it is no longer accurate in a complex sample context. Hence, there is a need for a modified prediction error estimator that is applicable to models based on, say, large-scale health surveys or political polling.  We now consider how to incorporate knowledge about the complex sampling design  giving rise to data for accurate estimation of $Err$. 

In the following, we make use of the \emph{superpopulation} framework \citep{horvitz1952generalization} for finite population analysis.  That is, we assume that the finite population $y_1\, \dots, y_N$ is generated independently (not necessarily identically) by the same mechanism $\g$, and that the data are then obtained via a (not necessarily uniform) sampling distribution, denoted $\pi$, where $\pi_i = Pr(y_i \in s)$, for sample $s$. Note that in prior sections we used $\g$ to denote the distribution producing the data, but we now use the same symbol to denote the distribution producing the finite population.  

If we know cov$_{\g}(\hat{\lambda}_i,y_i)$---or if we can obtain a consistent estimate of it---then Theorem \ref{optimism} provides an analytic, consistent estimator of $Err_i$ in the case of uniform sampling.  However, if the individual $y_i$s are collected according to a non-uniform sampling scheme with known or estimable sampling probabilities, it is still possible to obtain consistent estimates of generalization error by incorporating Horvitz-Thompson (HT) sampling weights into the error estimate. In order to address this issue, we must distinguish between different kinds of generalization error for the finite population framework. We now use $Err$ to denote  the finite population prediction error rate:
\begin{align*}
 Err = \frac{1}{N} \sum_{i=1}^N Q(y_i,\hat{\mu}_i) \ .
 \end{align*}
The superpopulation prediction error rate is the expected value of the finite population error rate $E_{\g} (Err)$.  Next, define the Horvitz--Thompson--Efron (HTE) estimator of the predictive error rate
 \begin{align*}
 \widehat{Err} = \frac{1}{N}\sum_{i=1}^n \frac{1}{\pi_i} \big(err_i + 2\ \mbox{cov}_{\g}(\hat{\lambda}_i,y_i)\big) \ ,
 \end{align*}
where  $\sum_{i=1}^n 1/\pi_i = N$, and $err_i$ is the same as in Equation (1). The following corollary follows easily from Theorem 2.1.

 \begin{corollary}{Corollary 3.3}{(Optimism theorem for biased samples)}
  The HTE estimator $\widehat{Err}$ is unbiased for the \emph{superpopulation} generalization error, i.e.
  \begin{align*}
      E(\widehat{Err}) = E_{\g}(Err) \, .
  \end{align*}
 \end{corollary}\label{cor}
\begin{proof}{Proof}{}
 Let $E_\pi$ denote expectation with respect to sampling mechanism. Then,
 \begin{align*}
 E(\widehat{Err}) &= E_{\pi,\g}\Big( \frac{1}{N}\sum_{i=1}^n \frac{1}{\pi_i} \big(err_i + 2\, \mbox{cov}_{\g}(\hat{\lambda}_i,y_i)\big) \Big)\\\nonumber
  &= E_{\g}\Big( \frac{1}{N}\sum_{i=1}^N \frac{1}{\pi_i} \,E_{\pi|\g} (I_{i\in s})\, \big(err_i + 2\, \mbox{cov}_{\g}(\hat{\lambda}_i,y_i)\big)\Big) \\ \nonumber
 &= E_{\g}\Big( \frac{1}{N}\sum_{i=1}^N \big(err_i + 2\, \mbox{cov}_{\g}(\hat{\lambda}_i,y_i)\big)\Big)  \nonumber
 = E_{\g}(Err)\, ,
  \end{align*}
  where the last equality results from Theorem \ref{optimism} and the linearity of the expectation operator.
 \end{proof}
 Thus, the HT extension (HTE) of the covariance penalty inflated estimator gives an unbiased estimator for the superpopulation prediction error irrespective of sample design. As a simple application of the law of large numbers, we know that $Err \stackrel{a.s.}{\rightarrow}E_{\g}(Err)$, and hence in the limit as $n,N \rightarrow \infty$ we have that $\hat{Err}$ is consistent for finite population generalization error $Err$.  For more details on asymptotics of the superpopulation framework see Fuller (2011, Section 1.3).
 
 \subsection{dAIC and the HTE}
 
 \cite{efron1986biased} shows that AIC and the covariance penalty inflated estimator of prediction error are asymptotically equivalent.  Here, we show the same is true for our HTE estimator and dAIC \citep{lumley2015aic}, a reweighted extension of AIC that accounts for complex sampling. We introduce dAIC in detail in the Appendix, and only provide a brief example here after proving the main result. 
 
The design based Akaike's information criterion is 
 \begin{align*}
 \mbox{dAIC} = -2 \hat{\ell}(\hat{\theta}) + 2\, \mbox{tr}\big\{ \hat{\mathcal{J}}(\hat{\theta})\hat{V}(\hat{\theta})\big\}\, ,
\end{align*}
where the first term is proportional to the HT reweighted likelihood
 \begin{align*}
 \hat{\ell}(\theta) = \frac{1}{N}\sum_{i=1}^n w_i  \, \ell_i(\theta) \, ,
 \end{align*}
 and the second term is the trace of the matrix product of 
 \begin{align*}
 \hat{\mathcal{J}}(\theta) = - \frac{1}{N}\sum_{i=1}^n w_i \frac{\partial^2 \ell_i (\theta)}{\partial\theta \partial \theta^T} \, ,
 \end{align*} 
 the HT reweighted log-likelihood Hessian and $\hat{V}(\hat{\theta})$, the regular `sandwich' estimator of the asymptotic covariance of the MLE.  The Appendix contains a thorough summary of dAIC and its terms, and Section \ref{example} provides a worked example.
 
 In the following theorem we establish the canonical result that, under non-uniform sampling, dAIC is a special case of the HTE estimator for standard generalized linear models (GLMs).

 \begin{theorem}{Theorem 3.1}{(Equivalence of dAIC and HTE)}
 The dAIC and HTE penalty terms correspond exactly, provided that:
 ($i$) a generalized linear model with canonical link is specified;
 ($ii$)  the weighted deviance loss is used and the model is fit by minimizing this loss function (which corresponds to maximizing the weighted log-likelihood).
 \end{theorem}\label{thm}
 
  \begin{proof}{Proof}{}
 Let $\lambda$ and $\mu$ denote the natural and mean parameters of the exponential family model
\begin{align*}
g_{\mu}(y) = \exp \big(\lambda \, y - \psi(\lambda) \big) \, .
\end{align*}
 Given observations $(y_i,x_i), i=1,\dots,n$, we adopt the GLM framework and assume that, for each observation, the canonical parameter is given by a linear combination of covariates: $\lambda_i = x_i^T\theta$. We show that the dAIC penalty is equal to the HTE penalty, i.e., that
 \begin{align*}
 \mbox{tr} \big(\hat{V}(\hat{\theta})\, \hat{\mathcal{J}}(\hat{\theta} ) \big) = \frac{1}{N} \sum_{i=1}^n \frac{1}{\pi_i} \widehat{\mbox{cov}}(\hat{\lambda}_i,y_i) \, ,
 \end{align*}
 when the estimate $\widehat{\mbox{cov}}$ is obtained using the analytic estimate (below) and \emph{not} obtained from the parametric bootstrap (although similar estimates are obtained in practice).
 We first use the following two facts about exponential family distributions to obtain the forms of $\hat{\mathcal{J}}(\hat{\theta})$ and $\hat{V}_U(\hat{\theta})$:
 \begin{align}\label{useful}
 \frac{\partial \psi}{\partial \lambda} = \mu  \quad \mbox{and} \quad \frac{\partial \mu}{\partial \lambda} = \frac{\partial^2 \psi}{\partial \lambda^2} = \mbox{cov}_\mu(y) \, .
 \end{align}
  Let $\Pi$ be the $n$ by $n$ diagonal matrix with $\Pi_{ii}=\pi_i$. $\hat{\mathcal{J}}(\hat{\theta})$ is defined as the HT weighted, observed Fisher information:
 \begin{align*}
 \hat{\mathcal{J}}(\hat{\theta})  &= - \frac{1}{N}\sum_{i=1}^n \frac{1}{\pi_i} \frac{\partial^2 \ell_i (\theta)}{\partial\theta \partial \theta^T}  \\ \nonumber
 &= \frac{1}{N}X^T (\Pi^{-1} \frac{\partial^2 \psi}{\partial \lambda^2}\big|_{\hat{\lambda}})\, X \\ \nonumber
 &= \frac{1}{N}X^T (\Pi^{-1} \hat{\Sigma}_M)\, X \, .
 \end{align*}
Here $\hat{\Sigma}_M=\Sigma(\hat{\mu})$ is a diagonal matrix with elements given by the model-based covariances $\Sigma(\hat{\mu})_{ii}=\mbox{cov}_{\hat{\mu}_i}(y_i)$.
Now, $\hat{V}_U(\hat{\theta})$ takes the form
\begin{align*}
 \frac{1}{N^2} X^T \Pi^{-1} \hat{\Sigma}_O\, \Pi^{-1} X \, ,
\end{align*}
for $\hat{\Sigma}_O$ a matrix of observed residuals with specific form depending on sample characteristics (see Appendix).

Next we need a formula for $\hat{\theta}$. Suppose that $\hat{\theta}$ is obtained by maximizing the weighted log-likelihood. Then $\hat{\theta}$ takes the form of the WLS solution
 \begin{align}\label{theta_hat}
 \hat{\theta} = (X^T(\hat{\Sigma}_M\Pi^{-1})X)^{-1} X^T(\hat{\Sigma}_M\Pi^{-1})z \, ,
 \end{align}
where $z$ is the linearization of the canonical link applied to $y$ called the `adjusted dependent variable' or the `working residual':
 \begin{align}\label{z}
 z = \hat{\lambda} + (y -\hat{\mu})\frac{\partial \lambda}{\partial \mu}\Big|_{\hat{\mu}} \, .
 \end{align}
Combining the formulas gives
\begin{align*}
\mbox{tr} \big(\hat{V}(\hat{\theta})\, \hat{\mathcal{J}}(\hat{\theta} ) \big) &= \mbox{tr} \big(\hat{V}_U(\hat{\theta})\, \hat{\mathcal{J}}(\hat{\theta} )^{-1} \big) \\ \nonumber
   &= \frac{1}{N}\mbox{tr} \Big( X^T(\Pi^{-1}\hat{\Sigma}_O\Pi^{-1})X\, \big(X^T(\hat{\Sigma}_M\Pi^{-1}) X\big)^{-1} \Big) \\ \nonumber
%   &=\frac{1}{N} \mbox{tr} \Big( \Pi^{-1}\,X \big(X^T(\hat{\Sigma}_M\Pi^{-1}) X\big)^{-1} X^T\Pi^{-1}\hat{\Sigma}_O \Big) \\ \nonumber
   &=\frac{1}{N} \mbox{tr} \Big( \Pi^{-1}\,X \big(X^T(\hat{\Sigma}_M\Pi^{-1}) X\big)^{-1} X^T\Pi^{-1}\hat{\Sigma}_M\hat{\Sigma}_M^{-1} \hat{\Sigma}_O \Big) \\ \nonumber
   &= \frac{1}{N}\mbox{tr} \Big( \Pi^{-1}\,X \big(X^T(\hat{\Sigma}_M\Pi^{-1}) X\big)^{-1} X^T\Pi^{-1}\hat{\Sigma}_M \,\frac{\partial \lambda}{\partial \mu}|_{\hat{\mu}}\, \hat{\Sigma}_O \Big) 
   \\ \nonumber
   &=\frac{1}{N} \mbox{tr} \Big( \Pi^{-1}\,X \big(X^T(\hat{\Sigma}_M\Pi^{-1}) X\big)^{-1} X^T\Pi^{-1}\hat{\Sigma}_M \,\widehat{\mbox{cov}}(z,y) \Big) \\ \nonumber
%   &= \frac{1}{N}\mbox{tr} \Big( \Pi^{-1} \widehat{\mbox{cov}}(X\hat{\theta},y) \Big) \\ \nonumber
   &= \frac{1}{N}\mbox{tr} \Big( \Pi^{-1} \widehat{\mbox{cov}}(\hat{\lambda},y) \Big) = \frac{1}{N} \sum_{i=1}^n \frac{1}{\pi_i} \widehat{\mbox{cov}}(\hat{\lambda}_i,y_i) \,,
\end{align*}
 completing the proof.
 \end{proof}

   \subsubsection{Example: linear regression model based on weighted independent sample}\label{example}
 Let $f_{\theta}(y|x)$ be the homoskedastic linear regression model with Gaussian errors and  regression coefficients $\theta$.  Then 
 \begin{align*}
 \hat{\theta} = (X^T\Pi^{-1}X)^{-1} X^T\Pi^{-1}Y\, ,
 \end{align*}
 and
  \begin{align}\label{J}
 \hat{\mathcal{J}}(\hat{\theta})= (X^T\Pi^{-1}X)/(N\hat{\sigma}^2)\, .
 \end{align}
 where $X$ is the $n\times p$ observed design matrix.
 Next, the estimated covariance matrix of $\hat{\theta}$ is given by
 \begin{align*}
 \hat{V}(\hat{\theta}) &= (N\hat{\sigma}^2)(X^T\Pi^{-1}X)^{-1} \times \\ \nonumber
 & \frac{X^T\Pi^{-1}\mbox{Diag}\big((Y-X\hat{\theta})(Y-X\hat{\theta})^T\big)\, \Pi^{-1}X}{(N\hat{\sigma}^2)^2}(X^T\Pi^{-1}X)^{-1}(N\hat{\sigma}^2) \\ \nonumber
 &= (X^T\Pi^{-1}X)^{-1}X^T\Pi^{-1}\hat{\Sigma}_O\, \Pi^{-1}X(X^T\Pi^{-1}X)^{-1}\, .
 \end{align*}
for $\hat{\Sigma}_O = \mbox{Diag}\big((Y-X\hat{\theta})(Y-X\hat{\theta})^T\big)$. The design-effect corrected penalty term is then given by
 \begin{align*}
  \mbox{tr}\big\{ \hat{\mathcal{J}}(\hat{\theta})\hat{V}(\hat{\theta})\big\} &= \mbox{tr}\big\{(X^T\Pi^{-1}X) \times  \\ \nonumber
  & (X^T\Pi^{-1}X)^{-1}X^T\Pi^{-1}\hat{\Sigma}_O\, \Pi^{-1}X(X^T\Pi^{-1}X)^{-1}\big\}/(N\hat{\sigma}^2) \\ \nonumber
 &= \mbox{tr}\big\{X^T\Pi^{-1}\hat{\Sigma}_O\, \Pi^{-1}X(X^T\Pi^{-1}X)^{-1} \big\}/(N\hat{\sigma}^2) \, .
 \end{align*}
 It follows that the dAIC for the classical linear regression case is given by
 \begin{align}\label{plusDelta}
 \frac{1}{N}\,\sum_{i=1}^n \frac{(y_i -\hat{\mu}_i)^2}{\pi_i\, \hat{\sigma}^2} + \frac{2}{N\hat{\sigma}^2}\, \mbox{tr}\big\{X^T\Pi^{-1}\hat{\Sigma}_O\, \Pi^{-1}X(X^T\Pi^{-1}X)^{-1} \big\}\, .
 \end{align}

 We now derive the HTE estimator and show it to be the same. Let
 \begin{align*}
err_i =  Q(y_i, \hat{\mu}_i)= (y_i - \hat{\mu}_i)^2
 \end{align*}
 be the loss function, then the inflation term for the HTE estimator is given by
 \begin{align*}
 \frac{1}{N} \sum_{i=1}^n \frac{2}{\pi_i}\, \widehat{\mbox{cov}}_{\g}(\hat{\mu}_i,y_i) &=  \frac{2}{N} \sum_{i=1}^n \frac{1}{\pi_i} \,\big(X(X^T\Pi^{-1}X)^{-1}X^T\Pi^{-1} \hat{\Sigma}_O\big)_{ii} \\ \nonumber
 &= \frac{2}{N} \mbox{tr} \big\{ \Pi^{-1}X(X^T\Pi^{-1}X)^{-1}X^T\Pi^{-1} \hat{\Sigma}_O \big\}\, .
 \end{align*}
 By the cyclic property of the trace, one has
 \begin{align}\label{plusCov}
 \hat{Err} &= \frac{1}{N}\,\sum_{i=1}^n \frac{1}{\pi_i} (y_i -\hat{\mu}_i)^2 + \frac{2}{N}\, \mbox{tr}\big\{X^T\Pi^{-1}\hat{\Sigma}_O\, \Pi^{-1}X(X^T\Pi^{-1}X)^{-1} \big\}
 \\ \nonumber
 &= \mbox{dAIC} \cdot \hat{\sigma}^2 \, .
 \end{align}
 Thus equivalence between the covariance penalty inflated prediction error estimator and dAIC clearly holds in this context.  This is expected as a special case of Theorem \ref{thm}.

\section{Simulation studies}\label{results}

\subsection{Consistency of the HTE estimator}

  \begin{table}
\tbl{Data generation schemes. For Scenario 1, the distribution of response $y_N^i$ is independent of the sampling probability $\pi^i$. In Scenario 2, the predictor influences the mean and variance of $y_N^i$, but is independent from the sampling probability. For Scenario 3, the distributions of $y_N^i$ and $\pi^i$ both depend on the index $i$. For Scenario 4, the distributions of $y_N^i$ and $\pi^i$ both depend on $X_N^i$. Since the variance of $y_N^i$ and the sampling probability both grow with $X_N^i$, one might expect optimism to be negative. This turns out to be the case (see Table 2).}{
\begin{tabular*}{30pc}{@{\hskip5pt}@{\extracolsep{\fill}}l@{}c@{}c@{}c@{}@{\hskip5pt}}
\toprule
  Scenario  & $X_N$ & $y_N$ & $\pi$  \\
 \colrule
1     & $X_N^i \stackrel{iid}{\sim}N(0,1)$ & $y_N^i \stackrel{ind}{\sim}N(X_N^i,1)$ & $\pi^i \propto \log(i)$  \\ 
\hline
2     & $X_N^i \stackrel{iid}{\sim}N(0,1)$ & $y_N^i \stackrel{ind}{\sim}N(X_N^i,|X_N^i|)$ & $\pi^i \propto \log(i)$ \\
      & $X_N^i \stackrel{iid}{\sim}N(0,1)$ & $y_N^i \stackrel{ind}{\sim}Bern \, \Phi(X_N^i)$ & $\pi^i \propto \log(i)$ \\
\hline
3     & $X_N^i \stackrel{iid}{\sim}N(0,1)$ & $y_N^i \stackrel{ind}{\sim}N(X_N^i,\log(i))$ & $\pi^i \propto \log(i)$ \\
\hline
4a     & $X_N^i \stackrel{iid}{\sim}N(0,1)$ & $y_N^i \stackrel{ind}{\sim}N(X_N^i,|X_N^i|)$ & $\pi^i \propto |X_N^i|$ \\
      & $X_N^i \stackrel{iid}{\sim}N(0,1)$ & $y_N^i \stackrel{ind}{\sim}Bern \, \Phi(X_N^i)$ & $\pi^i \propto |X_N^i|$ \\
      \cline{2-4}
4b     & $X_N^i \stackrel{iid}{\sim}N(0,1)$ & $y_N^i \stackrel{ind}{\sim}N(X_N^i,|X_N^i|)$ & $\pi^i \propto |X_N^i|^{-1}$ \\
      & $X_N^i \stackrel{iid}{\sim}N(0,1)$ & $y_N^i \stackrel{ind}{\sim}Bern \, \Phi(X_N^i)$ & $\pi^i \propto |X_N^i|^{-1}$ \\
  \botrule
\end{tabular*}
 }
 \label{table1}
\end{table}

    \begin{table}\label{table2}
\tbl{Population optimism versus HTE estimates. For both the finite population based optimism `Err-err' and the HTE estimated optimism $\hat{\Omega}$, means, medians, and empirical intervals based on 10,000 independent simulations are shown. Based on their mutual consistency for the superpopulation optimism, one expects the empirical means of the finite population optimisms to be close to the HTE estimates. Similar to other prediction error estimation methods, the HTE conditions on the observed support of $X$, and is inaccurate for Scenario 4b (see text).}{
\begin{tabular*}{30pc}{@{\hskip5pt}@{\extracolsep{\fill}}l@{}c@{}c@{}c@{}c@{}c@{}@{\hskip5pt}}  
\toprule
   & &  \multicolumn{2}{c}{Err-err}    &  \multicolumn{2}{c}{$\hat{\Omega}$}   \\
    \cline{3-6}
  & Scenario &  mean \{median\}&  $95\%$ Interval            & mean \{median\} & $95\%$ Interval     \\
 \colrule
Gaussian&1  & $~0.004$ \{0.004\}       & $(-0.087,  ~0.091)$   & $0.004$ \{0.004\} & $(~~0.003, 0.005)$ \\
&2   &  $~0.002$ \{0.007\}  & $(-0.192, ~0.168)$   &   $0.004$ \{0.004\}   &   $(-0.019,  0.026)$   \\
&3  & $~0.392$ \{0.524\}   & $(-9.792, 10.114)$  & $0.451$ \{0.449\} & $(~~0.360, 0.555)$\\
&4a    & $~0.023$ \{0.068\} & $(-0.674, ~0.446)$  & $0.009$ \{0.008\}    & $(-0.004,  0.024)$        \\
&4b    & $-0.007$ \{-0.008\} & $(-0.164, ~0.156)$  & $0.052$ \{0.036\}    & $(~~0.014, 0.189)$        \\
  \hline
 Bernoulli&2  & $~0.001$ \{0.001\} & $(-0.027,  ~0.028)$ & $0.001$ \{0.001\} &  $(~~0.000, 0.001)$\\
&4a  & $~0.015$ \{0.015\}    &  $(-0.041, ~0.065)$  &  $0.010$ \{0.007\}   &  (~~$0.002, 0.038)$    \\ 
&4b  & $-0.001$ \{-0.001\}    &  $(-0.050,  ~0.044)$  &  $0.000$ \{0.000\}   &  ($-0.000,  0.001)$    \\ 
  \botrule
\end{tabular*}}
\end{table}

In this section we illustrate the properties of the HTE estimator via Monte Carlo simulation under four potential non-uniform sampling designs and consider performance under a linear regression and logistic regression model fit. The simulated experiments encompass four simplified scenarios in which the relationship between model covariates, sampling probabilities, and model noise are allowed to differ. In most scenarios, the HTE estimator is shown to provide a useful estimate for the generalization error. We also show that the HTE estimator fails when sampling probabilities and model errors are strongly correlated.

In all four scenarios, the finite population is first generated and then subsampled.  The exact distributions of the data are given in Table \ref{table1}. In Scenario 1, the non-uniform sampling mechanism is independent of $X_N^i$ and $y_N^i$. For Scenario 2, both the mean and variance are functions of the predictor, but data generation is independent of the sampling mechanism.     In Scenario 3, the model errors and the sampling probabilities are a function of the same random variable $z^i$; in simulations $z_i$ was degenerate at the logarithm of index $i$. In Scenario 4, the model errors and the sampling probabilities are both dependent on the absolute value of $X_N^i$.
 
For each scenario, 10,000 simulations are run. Within each simulation, a finite population of size $N=100,000$ is generated from which a sample of size $n=1,000$ is obtained.  Then, weighted least squares regression or weighted logistic regression (using inverse probabilities as weights) are performed. In-sample error, extra-sample error, and the HTE estimator are recorded.  Thus, the simulations provide a 10,000 large empirical sample of the difference between extra-sample and in-sample errors and the HTE estimator. Results are shown in Table 2.

In the first column of Table 2, the mean, median,  and $0.025$ and $0.975$ quantiles of the simulated true optimism are presented. Note that in each scenario the optimism varies greatly between each of the 10,000 simulation iterations. The variance of the HTE estimates is much smaller in comparison, and, in most cases, their means and medians are extremely close to the means and medians of the empirical optimisms. The HTE estimator is not directly consistent for the population optimism but \emph{is} consistent for the \emph{superpopulation} optimism.  We therefore expect that the means and medians of Table 2 should become arbitrarily close as the number of simulation iterations gets large. 

Here we address the performance of the HTE estimator for Scenario 4. It is important to note  that, in covariance penalty inflated prediction error estimation, the estimate is not just \emph{based} on the observed data, but \emph{explicitly for} the error rate of future observations over the exact same support as the observed data. For the covariance penalty inflated prediction error methodology, it is easy to forget that the accuracy of the error estimate deteriorates as the distance (given by some metric on the data space) between observations and  future observations grows. This is another way of saying that error is a function of---among other things---the amount by which we use our error estimate inappropriately, i.e. as an estimate of something it is not  an explicit estimator of.  This fact explains why sampling probabilities that are strongly correlated with model errors would cause ostensibly inaccurate results (as in Table 2). Strong correlations between the sampling mechanism and model errors are problematic for the proposed methodology (and for dAIC) insofar as we choose to use it to generalize to future observations with drastically different support from that of the data observed.  Despite these facts, the estimator performs adequately in Scenario 4a. However, Scenario 4b is simulated with unusually perverse dependencies between sampling probabilities and model errors, enforcing a negative optimism that `breaks' our methodology. 

\subsection{Performance of the HTE estimator over AIC}

\begin{figure}
\includegraphics[width=\textwidth]{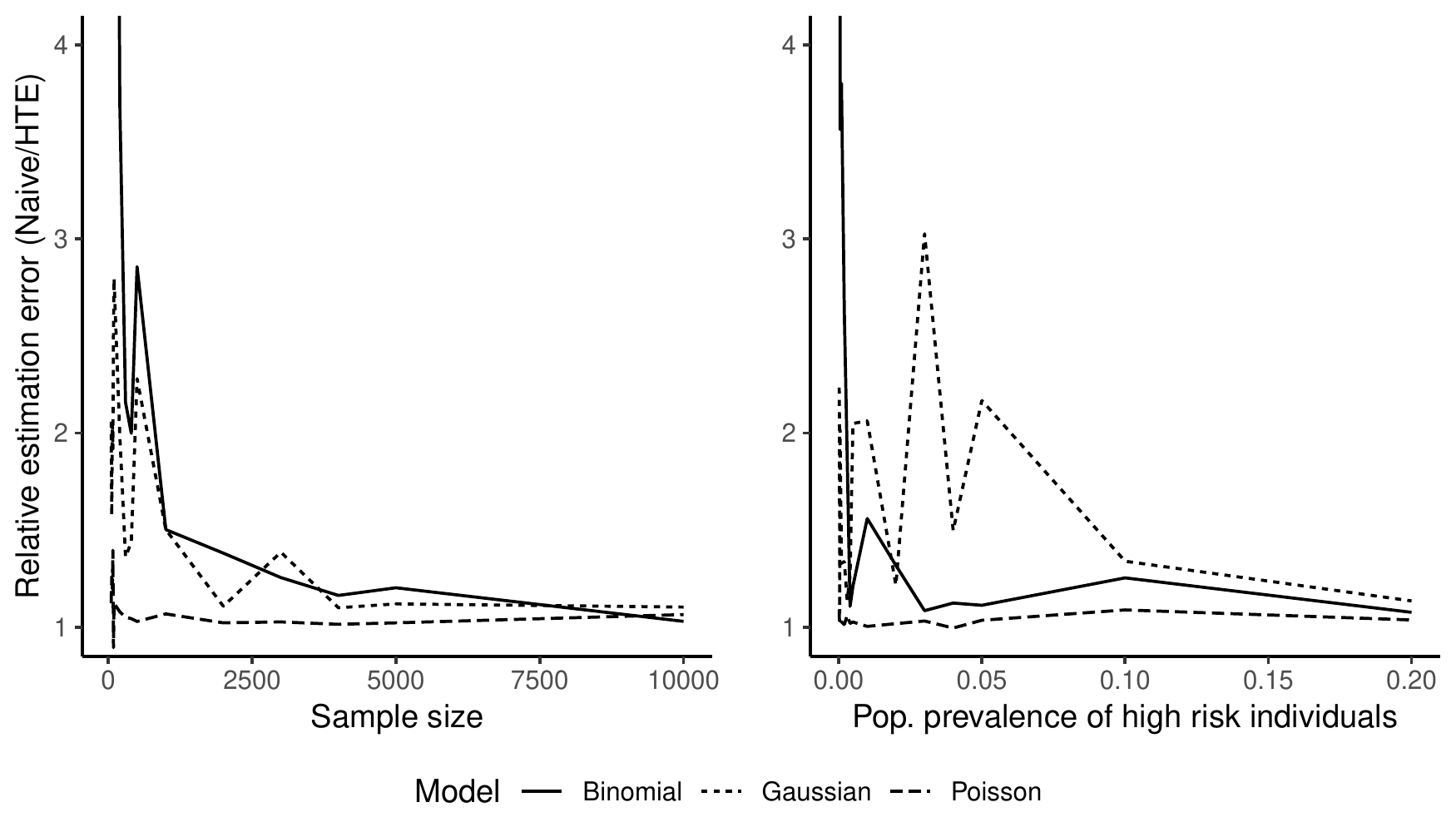}
\caption{Relative performance of HTE to `naive' AIC. For both plots, vertical axis is ratio between prediction error estimators \emph{estimation} error for true prediction error. Each setting is simulated independently 100 times and means are communicated. Right plot varies \emph{population} prevalence of high risk individuals keeping sampling proportions constant.  HTE is particularly beneficial for small samples and efficient designs.}
\end{figure}

Here we demonstrate the proposed HTE estimator's properties as a function of sample size and design efficiency in comparison to naive approaches that assume uniform sampling.  We simulate a situation in which high risk individuals are oversampled.  For binary outcome, this corresponds to a case-control study.   We compare the \emph{estimation} error of the \emph{prediction} error estimator to that of AIC for: 1) different sample sizes ($50$ through $10,000$) keeping population prevalence of high risk individuals ($1$ in $200$) and ratio ($20\%$) of high risk to normal individuals fixed; and 2) different population prevalence of high risk individuals ($1$ in $1,000$ to $1$ in $10$) keeping sample size ($1,000$) and ratio (again, $20\%$) of high risk to normal individuals fixed.

For each setting, real, count, and binary data are simulated from Gaussian, Poisson, and Bernoulli distributions, respectively, using the same systematic component with respective canonical links. For the Gaussian and Poisson scenarios, ``cases" are marked using a dummy variable indicating the top $.005$-quantile of the observed linear predictor values. Linear, logistic, and log-linear models are used for prediction.  True prediction error is obtained by applying the prediction rules to the rest of the finite population.  Each setting is simulated independently 100 times, and means are communicated.

Figure 1 exhibits the results in terms of the ratio of estimated prediction errors to the true prediction error.  This ratio is labeled the `relative estimation error'.  In both plots, the ratio in accuracies is most pronounced for the binary data case and more pronounced for the Gaussian data case than the count data case. For small samples, HTE outperforms AIC. As sample sizes grow, the difference between in-sample prediction error and extra-sample prediction error grows small, so the estimators  converge. When prevalence of high risk individuals in the population is small, the study design is more efficient and the HTE outperforms AIC.

\section{Prediction of renal function using data from the National Health and Nutrition Examination Survey}\label{NHANES}

  \begin{table}
\tbl{Comparing dAIC and HTE for prediction of GFR $(<60)$ with different covariates included. Results come from a logistic regression model using deviance loss. For each method, $\hat{p}$ is the estimated effective number of parameters. The HTE term $\hat{p}=n\,\hat{\Omega}/2$ is calculated using parametric bootstrap and presented as median from 100 simulations and 95\% empirical interval. Penalties increase with the number of covariates but are too small to influence generalization error estimates because of the large sample size.   Importantly, dAIC and HTE give similar results and have penalties that are larger that the usual AIC penalty. }{
\begin{tabular*}{30pc}{@{\hskip5pt}@{\extracolsep{\fill}}l@{}c@{}c@{}c@{}c@{}@{\hskip5pt}} 
\toprule
  Model      & Weighted deviance &$p$             &dAIC $\hat{p}$           & HTE $\hat{p}$      \\
 \colrule
 Age            & 9017.07 & 2 & 3.652    & 3.442 (2.550, 4.470) \\
+ BMI           & 8929.49 & 3 & 5.472   & 5.180  (3.998, 6.463)\\
+ Gender        & 8748.63 & 4 & 7.163    &  6.385 (5.339, 7.870) \\
+ Race/ethnicity& 8695.96 & 5 & 7.121    &     7.158 (6.036, 8.782)                    \\
%+ Smoke         & 8690.95 & 6 & 9.295    &      (, )                    \\
 \botrule
\end{tabular*}}
\end{table}

  \begin{table}
\tbl{Estimated error for k-nearest neighbours (kNN) classification. Again, GFR ($<$60) is being predicted, but errors are based on 0-1 loss.  As the number of voting neighbours increases, in-sample error increases, and the HTE optimism estimate $\hat{\Omega}$ (double the covariance penalty) decreases. Generalization error appears to hit its lowest point at around 30-NN.}{
\begin{tabular*}{30pc}{@{\hskip5pt}@{\extracolsep{\fill}}l@{}c@{}c@{}c@{}@{\hskip5pt}}
 \toprule
  Predictive model  & err & $\hat{\Omega}/2$ &$\hat{Err}$  \\
 \colrule
10-NN                  & 0.108 & 0.008 & 0.125 \\ 
20-NN                  & 0.113 & 0.004 & 0.121 \\ 
30-NN                  & 0.114 & 0.003 & 0.120 \\ 
40-NN                  & 0.116 & 0.002 & 0.121 \\ 
  \botrule
\end{tabular*}}
\end{table}

We consider data from the Third National Health and Nutrition Examination Survey (NHANES III) with the illustrative goal of constructing a model for predicting abnormal renal function as proxied by a estimated glomerular filtration rate (GFR) of less than 60 millilitres per minute per 1.73m$^2$. Briefly, NHANES III was one of several periodic surveys conducted by the National Center for Health Statistics (NCHS). The survey was conducted during 1988 to 1994, and was designed to provide national estimates of health and nutritional status in the civilian non-institutionalized United States population aged 2 months and older. Children ages two months to five years, persons 60 years and older, Mexican-American persons, and non-Hispanic black persons were sampled at rates substantially higher than their proportions in the general population \cite{nhanes3}. To estimate renal function, we use the Modification of Diet in Renal Disease (MDRD) equation for GFR based on serum levels and demographic covariates \cite{levy1999green}. Specifically, GFR was estimated as
\begin{align*}
    \mbox{GFR}_{\mbox{MDRD}}&=170\times\mbox{SCr}^{-0.999}\times \mbox{ageyrs}^{-0.176} \times \mbox{BUN}^{-0.170}  \times \\ \nonumber
    &\mbox{Salb}^{0.318} \times 1.180^{\mbox{black}} \times 0.762^{\mbox{female}} \, ,
\end{align*} 
where SCr denotes serum creatinine, BUN denotes blood urea nitrogen, Salb denotes serum albumin, and black and female denote indicators of non-Hispanic black race and sex, respectively. \cite{coresh2002calibration} have previously reported that the assay used for measuring serum creatinine in the NHANES study resulted in creatinine levels systematically higher than those used to obtain the MDRD prediction model. As a consequence, they suggest creatinine values from NHANES III be recalibrated to account for an average overestimate of 0.23 mg/dL. All analyses presented here have performed the recommended recalibration.

Grade 3 chronic kidney disease (CKD) is defined as a GFR less than 60 millilitres per minute per 1.73m$^2$ and is associated with increased morbidity and risk of end stage renal disease. As such we consider building a model for predicting grade 3 CKD. To illustrate the generalizability of the proposed HTE estimator, we approach this binary prediction task from two separate vantage points and using two different loss functions.  First, we use logistic regression to demonstrate the empirical equivalence between dAIC and the HTE approach; here, deviance loss is used. Second, we compare a number of prediction models with respect to in-sample and estimated extra-sample errors; here, 0-1 loss is used. Using the NHANES data, we first demonstrate that dAIC and HTE give similar results for a binary prediction task with deviance loss. Second, using 0-1 loss, we calculate the in-sample error and HTE estimator for the k-neareast neighbours method (for which dAIC is not available) and show how the HTE estimate might influence prediction model preference.  Data and code for all empirical studies has been included in the supplement.

\subsection{dAIC and the HTE}
Deviance loss is used for the dAIC/HTE comparison, so $\hat{\lambda}_i$ is given by the log-odds. Table 3 shows the results from the dAIC/HTE comparison based on a logistic regression model with different covariate combinations. $p/n$, the number of covariates divided by sample size, is used as a reference that accords with a scaled traditional AIC, where the scaling is to meant for easy comparison to prediction error. dAIC is similarly scaled.  In general, all optimism estimates grow with the model size, but all generalization error estimates get smaller with model size. Indeed, the optimism estimates are kept small by the large size of the data sample. We note that dAIC and HTE are close. 

\subsubsection{Estimating the covariance penalty}
The covariance between the linear predictor $\hat{\lambda}_i$ and the outcome $y_i$ is rarely known outside of a few basic examples and approximations.  As a result, \cite{efron1986biased} suggested the use of the parametric bootstrap to produce empirical covariances between simulated outcomes and their resulting fitted values.   In this comparison, we use a stratified quasi-binomial GLM, where `quasi' denotes a shared intra-primary sampling unit (PSU) dispersion parameter $\phi$ satisfying:
\begin{align*}
\mbox{var}(y_i)= p_i(1-p_i)\phi \quad \mbox{and} \quad \phi=(1 + (n_j-1)\rho) \, ,
\end{align*}
for $n_j$ the size of PSU $j$ and $\rho$ the within-PSU correlation shared across all PSUs. If $\hat{\rho}$ differs from zero, then $\hat{\phi}$ should be multiplied by the naive parametric bootstrap estimated covariance
\begin{align*}
\sum_{b=1}^B \hat{\lambda}^{*b}_i (y_i^{*b}-y_i^{*\cdot}) \, ,
\end{align*}
making the correct estimate
\begin{align*}
\widehat{\mbox{cov}}_i = \hat{\phi}\sum_{b=1}^B \hat{\lambda}^{*b}_i (y_i^{*b}-y_i^{*\cdot}) \, .
\end{align*}
That said, this point is moot in a number of ways. First, when using a GLM (as is the case in this example), it is less computationally intensive to use dAIC, the fully analytic special case of the HTE.  Second, in a study such as NHANES with PSUs on the order of $600$, within-PSU correlations tend to be small (in this paper, $|\hat{\rho}|< 5\times 10^{-4}$. Third, survey structure is often approximated for end-users.  Fourth, if the purpose is model comparison, multiplying by a scalar effects each fit equally.  Nonetheless, Table 3 presents dAIC (using sandwich estimator) and HTE (using parametric bootstrap) side-by-side for the to show that results are similar as is to be expected.

\subsection{k-Nearest neighbours}

We also compare the in-sample error and HTE estimated errors for k-nearest neighbours (kNN) models with different values of k. Here we adopt 0-1 loss, so $\hat{\lambda}_i$ is given by -1 for $p_i<.5$ and 1 for $p_i\geq.5$ \citep{efron2004estimation}. Results are shown in Table 4.  In general, the larger k is, the smoother the decision rule. In terms of the bias-variance trade-off, this amounts to more bias and less variance.  Indeed, as k increases from 10 to 40, in-sample prediction error increases, but the covariance-inflation decreases.  The HTE estimator appears to achieve an optimum somewhere around $k=30$.

\section{Discussion}

Motivated by the increasing importance of algorithmic prediction methodologies and the need to effectively make predictions in the public health and medical sectors, we present a prediction error estimation methodology with the hope that it will provide for the rigorous comparison of competing prediction rules obtained from complex survey data.  We show that our Horvitz--Thompson--Efron (HTE) estimator is accurate and somewhat robust for GLMs and algorithmic prediction methods. Moreover, we prove that the HTE generalizes dAIC (an AIC variant for complex samples) in the exact same way that Efron's covariance penalty inflated estimator generalizes AIC.  We empirically demonstrate this fact via simulation and also by considering the prediction of chronic kidney disease using data from NHANES III, a large public health survey with prescribed sampling weights.

There is a trend in medicine toward increasingly personalized treatment.  Such treatment is essentially a prediction task and, as such, is subject to the bias-variance trade-off. To help avoid over-fitting when training the necessary predictive models, it will be necessary to use large swaths of public health data, the majority of which arises from complex sampling procedures.  We therefore expect that our proposed methodology and its extensions will become increasingly important for model scoring in the context of personalized medicine.  Causal inference from observational data is in some ways the opposite challenge of personalized medicine, although the two are closely tied together.  Moreover, methods in observational causal inference often make use of the Horvitz--Thompson reweighting procedure.  We are particularly interested in the question of whether the proposed HTE estimator may be extended using reweighting procedures commonly used in causal inference and whether this methodology might be useful for effective personalized medicine.

We know of two immediate extensions to the methodology proposed here.  Whereas the HTE estimator is based on the closed-form model optimism, algorithmic and non-analytic prediction error estimators are more common in the machine learning literature.  We therefore anticipate the extension of both cross-validation and bootstrap prediction error estimation to the complex sample domain. 

\section*{ACKNOWLEDGEMENTS}

Research reported in this article was supported by the National Institutes of Health under award number R01AG053555. %(DLG).

\clearpage
 \appendix
 
 \subsection{Design-based AIC}\label{dAIC}

 \cite{lumley2015aic} proposed an extension to AIC under a non-uniform sampling regime by adopting the above superpopulation framework. We do not know the true distribution $\g$, and we seek to minimize the KL divergence between a plausible conditional distribution $f_{\theta}(y|\mathbf{x})$ and $g(y)$ for observed covariate vector $\mathbf{x}$.  As in the uniform sampling case, this corresponds to maximizing the  log-likelihood $\ell (\theta)$ of the assumed model.  For complex samples, this may be estimated using Horvitz--Thompson \citep{horvitz1952generalization} weights:
 \begin{align*}
 \hat{\ell}(\theta) = \frac{1}{N}\sum_{i=1}^n w_i  \, \ell_i(\theta) \, ,
 \end{align*}
 where $n$ is the size of $s$, $N$ is the size of the finite population, $w_i \propto 1/\pi_i$, and $\sum_{i=1}^n w_i = N$. In general, the weights need not be the true sampling weights and may be adjusted to account for nonresponse or calibrated toward population totals. Weight estimation methodologies are well understood. Important examples in the biostatistics literature include \cite{robins1994estimation} and \cite{seaman2013review}. Examples from the survey sampling literature include \cite{valliant1993poststratification} and \cite{deville1992calibration}, and \cite{lumley2011connections} consider connections between the areas.
 
 Now, suppose that $\theta^*$ and $\hat{\theta}$ are obtained by solving the score equation and pseudo-score equation, respectively:
 \begin{align*}
 U(\theta)= \frac{\partial \ell(\theta)}{\partial \theta} = 0\,, \quad \mbox{and} \quad \hat{U}(\theta)= \frac{\partial \hat{\ell}(\theta)}{\partial \theta} = 0 \, .
 \end{align*}
In the context of AIC, we are interested in estimating $E_{\g} (\ell (\hat{\theta}))$, the expected value of the log-likelihood $\log f_{\theta} (y|\mathbf{x})$ evaluated at $\hat{\theta}$ with respect to the true superpopulation distribution $\g$. Then it is shown in \cite{lumley2015aic} that
 \begin{align}\label{result_lumley}
 E_{\g}\big(\hat{\ell}(\hat{\theta}) \big) = E_{\g} \big(\ell (\hat{\theta})\big) + \frac{1}{n}\mbox{tr}(\Delta) + o_p(n^{-1}) \, 
 \end{align}
 where $\Delta = I(\theta^*)V(\theta^*)$, $V(\theta^*)$ is the asymptotic covariance of $\sqrt{n} \hat{\theta}$, and 
 \begin{align*}
 \mathcal{I}(\theta) = E_{\pi,\g}\big(\hat{\mathcal{J}}(\theta)\big) = - E_{\g} \Big(\frac{\partial^2 \ell (\theta)}{\partial\theta \partial \theta^T}     \Big) \, ,
 \end{align*}
 for
 \begin{align*}
 \hat{\mathcal{J}}(\theta) = - \frac{1}{N}\sum_{i=1}^n w_i \frac{\partial^2 \ell_i (\theta)}{\partial\theta \partial \theta^T} \, .
 \end{align*}
 Note that $\mathcal{I}$ is just the Fisher information corresponding to distribution $\g$ and that $\hat{\mathcal{J}}(\hat{\theta})$ reduces to the observed Fisher information when the sample is collected uniformly. 
 Equation \eqref{result_lumley} results in a design-based formulation of AIC for complex data, dAIC:
 \begin{align*}
 \mbox{dAIC} = -2 \hat{\ell}(\hat{\theta}) + 2\, \mbox{tr}\big\{ \hat{\mathcal{J}}(\hat{\theta})\hat{V}(\hat{\theta})\big\}\, ,
 \end{align*}
where $\hat{V}(\hat{\theta})$ is the sandwich estimator for $V(\theta^*)$:
\begin{align}\label{sandwich}
\hat{V}(\hat{\theta}) = \hat{\mathcal{J}}(\hat{\theta})^{-1} \hat{V}_U(\hat{\theta}) \hat{\mathcal{J}}(\hat{\theta})^{-1}
\end{align}
for $\hat{V}_U(\hat{\theta})$ a consistent estimator of cov$\big(\sqrt{n} \hat{U} (\theta)\big)$.  Thus dAIC may be rewritten as 
\begin{align*}
 \mbox{dAIC} = -2 \hat{\ell}(\hat{\theta}) + 2\, \mbox{tr}\big\{ \hat{\mathcal{J}}(\hat{\theta})^{-1}\hat{V}_U(\hat{\theta})\big\}\, .
\end{align*}
When the weights are uniform, dAIC reduces to a robust version of AIC called Takeuchi's information criterion \cite{takeuchi1976distribution}. If, in addition, the model is correctly specified, dAIC reduces to AIC \cite{lumley2015aic}.

 \paragraph*{The meat of the sandwich}\label{meat}
 
 The sandwich estimator for $V(\theta^*)$ is provided in Equation \eqref{sandwich}. The meat of this sandwich  is the estimated asymptotic covariance $\hat{V}_U(\hat{\theta})$ of the score function $\hat{U}(\hat{\theta})$. In its most general form, we have
 \begin{align*}
 \hat{V}_U(\hat{\theta}) = \hat{U}(\hat{\theta})\hat{U}(\hat{\theta})^T \, .
 \end{align*}
 Among other things, the example of Section \ref{example} shows that for linear regression in the non-uniform, unstratified sampling case, $\hat{V}_U(\hat{\theta})$ takes the form
 \begin{align*}
 \frac{X^T\Pi^{-1}\hat{\Sigma}_O\, \Pi^{-1}X}{(N\hat{\sigma}^2)^2}\, ,
 \end{align*}
 where $\hat{\Sigma}_O$ is the diagonal matrix of pointwise residuals. In general, for exponential family GLMs for non-uniform, unstratified samples, we have
  \begin{align*}
\hat{V}_U(\hat{\theta}) = \frac{1}{N^2} X^T\Pi^{-1}\hat{\Sigma}_O\, \Pi^{-1}X\, ,
 \end{align*}
 for $\hat{\Sigma}_O$ the diagonal matrix of observed residuals.
    The NHANES data considered in Section \ref{NHANES} is obtained from a stratified sample with intra-stratum and inter-PSU correlations as well as intra-PSU correlations.  In such a sample with strata $h=1,\dots,H$, $\hat{V}_U(\hat{\theta})$ takes the form 
   \begin{align*}
\hat{V}_U(\hat{\theta}) = \frac{1}{N^2} X^T\Pi^{-1} bdiag \big(\hat{\Sigma}_O^1,\dots,\hat{\Sigma}_O^H  \big) \Pi^{-1}X\, ,
 \end{align*}
 assuming observations are ordered according to stratum membership. Here, \emph{bdiag} indicates a block-diagonal structure, and $\hat{\Sigma}_O^h$ itself has a block structure corresponding to individual PSUs:
\[ \hat{\Sigma}_O^h =  \left( \begin{array}{llc}
\hat{\Sigma}_{1}^h & \hat{\Sigma}_{12}^h & \dots \\
\vdots & \ddots &  \\
\hat{\Sigma}_{1n_h}^h &  & \hat{\Sigma}_{n_h}^h \end{array} \right)\, .\] 
 In this formula, blocks along the diagonal are given by 
 \begin{align*}
 \hat{\Sigma}_j^h = (Y^h_j-\hat{\mu}^h_j)(Y^h_j-\hat{\mu}^h_j)^T = \sum_{i,i'\in j} (y^h_i-\hat{\mu}^h_i)(y^h_{i'}-\hat{\mu}^h_{i'}) \, ,
 \end{align*}
 where $i,i'$ denote individuals and $j$ denotes the PSU. Off-diagonal blocks take the form
 \begin{align*}
 \hat{\Sigma}^h_{jj'} = (Y^h_j-\hat{\bar{\mu}}^h_j)(Y^h_{j'}-\hat{\bar{\mu}}^h_{j'})^T = \sum_{i\in j,\, i' \in j'} (y^h_i-\hat{\bar{\mu}}^h_j)(y^h_{i'}-\hat{\bar{\mu}}^h_{j'}) \, 
 \end{align*}
 where $\hat{\bar{\mu}}^h_j$ is the average predicted value for PSU $j$, and $\hat{\bar{\mu}}^h_j$ is this average multiplied by an $n_j$-vector of ones.  The upshot is that in the above (the proof of Theorem 3.1, in particular), we can ignore the particular sampling structure and simply write
 \begin{align*}
 \hat{V}_U(\hat{\theta}) = \frac{1}{N^2} X^T\Pi^{-1} \hat{\Sigma}_O   \Pi^{-1}X\, ,
 \end{align*}
 letting $\hat{\Sigma}_O$ take whichever form appropriate for whichever sampling structure.

 \CJShistory
\end{document}